\documentclass[conference]{IEEEtran}
\usepackage{graphicx}
\usepackage{booktabs}
\usepackage{multirow}
\usepackage{amsmath}
\usepackage{url}
\usepackage{hyperref}
\usepackage{xcolor}
\usepackage{cite}

\title{From BM25 to Corrective RAG: Benchmarking Retrieval Strategies for Text-and-Table Documents}

\author{
  \IEEEauthorblockN{Meftun Akarsu}
  \IEEEauthorblockA{Technische Hochschule Ingolstadt\\mea5963@thi.de}
  \and
  \IEEEauthorblockN{Recep Kaan Karaman}
  \IEEEauthorblockA{Uludag University\\kaankaraman@uludag.edu.tr}
  \and
  \IEEEauthorblockN{Christopher Mierbach}
  \IEEEauthorblockA{Radiate\\christopher@radiate.com}
}

\begin{document}
\maketitle

\begin{abstract}
Retrieval-Augmented Generation (RAG) systems critically depend on retrieval quality, yet no systematic comparison of modern retrieval methods exists for heterogeneous documents containing both text and tabular data. We benchmark ten retrieval strategies spanning sparse, dense, hybrid fusion, cross-encoder reranking, query expansion, index augmentation, and adaptive retrieval on a challenging financial QA benchmark of 23,088 queries over 7,318 documents with mixed text-and-table content. We evaluate retrieval quality via Recall@$k$, MRR, and nDCG, and end-to-end generation quality via Number Match, with paired bootstrap significance testing. Our results show that (1)~a two-stage pipeline combining hybrid retrieval with neural reranking achieves Recall@5 of 0.816 and MRR@3 of 0.605, outperforming all single-stage methods by a large margin; (2)~BM25 outperforms state-of-the-art dense retrieval on financial documents, challenging the common assumption that semantic search universally dominates; and (3)~query expansion methods (HyDE, multi-query) and adaptive retrieval provide limited benefit for precise numerical queries, while contextual retrieval yields consistent gains. We provide ablation studies on fusion methods and reranker depth, actionable cost-accuracy recommendations, and release our full benchmark code.
\end{abstract}


\section{Introduction}
\label{sec:intro}

The introduction of the Transformer architecture \cite{vaswani2017attention} and subsequently pretrained language models \cite{devlin2019bert} fundamentally changed how machines process and reason over text. Yet even the most capable language models face a hard limit: their knowledge is frozen at training time. Retrieval-Augmented Generation (RAG) \cite{lewis2020rag} addresses this by coupling a language model with a retrieval component that fetches relevant documents at inference time, grounding generation in external evidence.

The retrieval component is the most critical part of any RAG system. A language model cannot reason over documents it never receives. Despite this, retrieval receives far less systematic attention than generation. Dozens of retrieval methods exist, benchmarks are inconsistent, and there is little guidance on which method works best for a given document type.

The problem is harder for \textbf{documents that mix text and tables}. Financial filings, earnings reports, and regulatory documents are structured this way: text provides context while tables carry the precise figures. A question like ``What was the year-over-year revenue growth?'' requires locating both the right document and the right cell within it. Semantic retrieval methods miss exact numerical targets. Lexical methods miss paraphrased context. Neither alone is sufficient.

We study this through systematic benchmarking. Using a financial QA benchmark of 23,088 queries over 7,318 documents with mixed text-and-table content, we evaluate retrieval methods ranging from classical sparse retrieval to hybrid fusion pipelines and neural rerankers. The original benchmark paper tested only six retrieval methods with two metrics. Many approaches that matter in practice, including contextual retrieval, CRAG, and modern reranking models, have never been evaluated in this setting.

\paragraph{Contributions.} This paper makes four contributions:
\begin{enumerate}
    \item A systematic benchmark of ten retrieval methods on a text-and-table financial QA corpus, the most comprehensive evaluation of its kind.
    \item Multi-dimensional evaluation covering retrieval metrics (Recall@$k$, MRR, nDCG, MAP) and generation metrics (Number Match) with paired bootstrap significance testing.
    \item Ablation studies isolating the effects of fusion strategies and reranker candidate depth.
    \item Actionable recommendations for practitioners building RAG systems over heterogeneous documents, grounded in empirical cost-accuracy analysis.
\end{enumerate}

\section{Related Work}
\label{sec:related}

\paragraph{RAG Retrieval Methods.}
Classical sparse retrievers such as BM25 remain competitive baselines, particularly in zero-shot out-of-domain settings \cite{thakur2021beir}, due to their lexical precision. Learned sparse models like SPLADE \cite{formal2021splade} extend this by learning neural term expansion weights, and recent work shows that decoder-only LLM backbones further improve sparse retrieval quality \cite{doshi2024mistralsplade}. On the dense side, dual-encoder architectures pioneered by DPR \cite{karpukhin2020dpr} encode queries and passages into a shared embedding space. Subsequent models including E5 \cite{wang2022e5}, E5-Mistral \cite{wang2024e5mistral}, and BGE-M3 \cite{chen2024bgem3} have advanced the state of the art on MTEB \cite{muennighoff2023mteb} and MMTEB \cite{enevoldsen2025mmteb} through contrastive pretraining and instruction-tuned embedding generation. Late-interaction models such as ColBERTv2 \cite{santhanam2022colbertv2} and Jina-ColBERT-v2 \cite{sturua2024jinacolbert} retain per-token representations for fine-grained matching while still allowing document precomputation. Hybrid retrieval combines sparse and dense signals via Reciprocal Rank Fusion \cite{cormack2009rrf} or convex score combination, consistently improving recall by 15--30\% over single-method pipelines \cite{li2025ragbestpractices}. Two-stage reranking with cross-encoder models further refines ranked lists, with recent benchmarks reporting up to 28\% nDCG@10 improvement at modest latency cost \cite{gao2024ragsurvey}.

\paragraph{Query-Side Retrieval Strategies.}
Several methods improve retrieval by modifying the query rather than the index. HyDE \cite{gao2022hyde} generates a hypothetical answer document at query time and retrieves using its embedding rather than the original query. RAG-Fusion \cite{raudaschl2023ragfusion} issues multiple LLM-rewritten query variants and merges results via RRF. Both approaches aim to close the gap between short user queries and longer document passages, though their effectiveness depends heavily on whether the LLM can generate plausible pseudo-documents for the target domain.

\paragraph{Index-Side and Structural Retrieval Strategies.}
An alternative is to enrich document representations at indexing time. Contextual Retrieval \cite{anthropic2024contextual} prepends LLM-generated summaries to each chunk before indexing, while HyPE \cite{vake2025hype} precomputes hypothetical questions per chunk, transforming retrieval into question-to-question matching and improving context precision by up to 42 percentage points without added query-time cost. Late Chunking \cite{guenther2024latechunking} preserves cross-chunk context by applying long-context embeddings before splitting. At a higher level of abstraction, RAPTOR \cite{sarthi2024raptor} builds recursive tree-structured summaries for multi-granularity retrieval, and GraphRAG \cite{edge2024graphrag} constructs entity-relation graphs for query-focused summarization. On the adaptive front, Self-RAG \cite{asai2023selfrag} trains models to decide whether retrieval is needed at all, while CRAG \cite{yan2024crag} triggers corrective web searches when retrieved document quality is low.

\paragraph{Text-and-Table Question Answering.}
Answering questions over documents that contain both text and tables requires locating evidence across heterogeneous content types and often performing numerical reasoning. Chen et al.~\cite{chen2020hybridqa} introduced HybridQA, the first large-scale dataset requiring multi-hop reasoning over linked Wikipedia tables and passages. Chen et al.~\cite{chen2021ottqa} extended this to an open-domain setting in OTT-QA, retrieving from over 400K tables and 5M passages. In the financial domain, FinQA \cite{chen2021finqa} provides expert-annotated question-program pairs over earnings reports, TAT-QA \cite{zhu2021tatqa} focuses on numerical operations over hybrid tabular-textual contexts, and ConvFinQA \cite{chen2022convfinqa} extends FinQA to multi-turn reasoning. A recent survey of table QA in the LLM era \cite{nan2025tableqasurvey} finds that retrieval of the correct heterogeneous context remains the primary bottleneck, even as generation quality improves. Our work directly targets this bottleneck.

\paragraph{RAG Benchmarks and Evaluation.}
BEIR \cite{thakur2021beir} established zero-shot retrieval evaluation across 18 diverse datasets and remains the standard reference point for retrieval method comparison. KILT \cite{petroni2021kilt} unified five knowledge-intensive task types over a shared Wikipedia snapshot. More recent RAG-specific benchmarks include RGB \cite{chen2024rgb}, which tests robustness to noise and counterfactual context; CRAG \cite{yang2024crag}, which spans five domains with temporal dynamics; and RAGBench \cite{friel2024ragbench}, which provides 100K examples with the TRACe evaluation framework. RAGAS \cite{es2024ragas} and ARES \cite{saadfalcon2024ares} introduced automated evaluation without gold labels, enabling scalable faithfulness and relevance measurement. Li et al.~\cite{li2025ragbestpractices} provide a systematic study of RAG design choices at COLING 2025. None of these benchmarks focus on retrieval method comparison over documents with mixed text-and-table content. The benchmark we use \cite{strich2026t2ragbench} unifies FinQA, ConvFinQA, and TAT-DQA into 23,088 queries over 7,318 financial documents, and where the original paper tested six methods with two metrics, we evaluate ten methods with a comprehensive set of retrieval and generation metrics.

\section{Methodology}
\label{sec:method}

\subsection{Dataset}
\label{sec:dataset}

We evaluate on T\textsuperscript{2}-RAGBench \cite{strich2026t2ragbench}, a financial QA benchmark accepted at EACL 2026. The dataset contains 23,088 question-context-answer triples drawn from three source datasets: FinQA \cite{chen2021finqa} (8,281 pairs), ConvFinQA \cite{chen2022convfinqa} (3,458 pairs), and TAT-DQA \cite{zhu2021tatqa} (11,349 pairs), covering 7,318 unique financial documents averaging approximately 920 tokens each. Each document contains a mix of text and markdown-formatted tables extracted from real SEC filings and annual reports.

The benchmark's core design decision distinguishes it from prior financial QA datasets. FinQA, ConvFinQA, and TAT-DQA were originally constructed in an oracle-context setting, where the relevant document is provided directly to the model. Questions in that setting are context-dependent: the same question may have different correct answers depending on which document is supplied, making them unsuitable for evaluating retrieval. T\textsuperscript{2}-RAGBench addresses this by reformulating all questions using Llama 3.3-70B to incorporate identifying information such as company name, sector, and report year, producing questions with exactly one correct answer regardless of context. Human experts validated a random sample of 100 questions per subset: only 7.3\% of original questions were context-independent, compared to 83.9\% after reformulation, with an inter-annotator agreement of Cohen's $\kappa = 0.58$.

All answers are numerical. The original paper evaluated six retrieval methods using Number Match and MRR@3, finding that the best method (Hybrid BM25) reached only 41\% Number Match against an oracle-context ceiling of 72--79\%, a gap of more than 30 percentage points. We extend this evaluation to ten methods with a broader set of retrieval and generation metrics to systematically characterize where this gap comes from.

\subsection{Retrieval Methods}
\label{sec:methods}

\paragraph{BM25.}
We use Okapi BM25 \cite{robertson1994okapi} with $k_1 = 1.2$ and $b = 0.75$ via the \texttt{rank\_bm25} library. BM25 scores documents by weighted term-frequency overlap with sub-linear saturation and document-length normalization. The parameter $k_1$ controls term-frequency saturation and $b$ controls the degree of length normalization; the values $k_1 = 1.2$ and $b = 0.75$ are the canonical defaults from the original Okapi system. BM25 provides strong lexical matching for domain-specific terminology such as company names, financial metrics, and fiscal period identifiers that appear verbatim in both queries and documents, making it a competitive baseline on this corpus.

\paragraph{Dense Retrieval.}
We encode all queries and documents using OpenAI \texttt{text-embedding-3-large} (3,072 dimensions) via Azure AI Foundry. Document embeddings are indexed with FAISS \texttt{IndexFlatIP} for exact inner-product search, ensuring exhaustive nearest-neighbor retrieval with no approximation error. At query time the top-$k$ results are returned by cosine similarity. This configuration isolates the effect of the embedding model from any index approximation artifacts.

\paragraph{Hybrid Retrieval (RRF).}
Hybrid retrieval fuses the ranked lists of BM25 and dense retrieval via Reciprocal Rank Fusion \cite{cormack2009rrf}. For each document $d$ at rank $r_i(d)$ in retriever $i$, the fused score is:
\begin{equation}
\text{RRF}(d) = \sum_{i} \frac{1}{k + r_i(d)}
\end{equation}
with smoothing constant $k = 60$, the value used in the original paper. We retrieve full ranked lists from both methods, compute RRF scores over their union, and return the top-$k$ by fused score. RRF is unsupervised, requires no score normalization, and consistently outperforms individual retrievers and alternative fusion strategies such as Condorcet and CombMNZ \cite{cormack2009rrf}.

\paragraph{Hybrid + Cohere Rerank.}
We apply a two-stage pipeline: hybrid RRF retrieves 50 candidate documents, which are then reranked by Cohere Rerank v4.0 Pro \cite{cohere2025rerank4}, returning the top 10. Unlike bi-encoder models that encode queries and documents independently, cross-encoders process the query and each candidate jointly, producing query-aware relevance scores that capture semantic relationships pointwise retrieval cannot \cite{nogueira2019passage}. Cohere Rerank v4.0 Pro was benchmarked specifically on finance-domain retrieval tasks at release, making it well suited to this corpus. This configuration measures whether the added cost of a reranking stage produces meaningful gains on text-and-table documents.

\paragraph{HyDE.}
HyDE \cite{gao2022hyde} addresses the asymmetry between short queries and long documents by generating a hypothetical answer passage at query time and retrieving with its embedding rather than the original query embedding. The generated document may contain hallucinations, but the dense encoder grounds it to the actual corpus by mapping it into the same embedding space as real documents \cite{gao2022hyde}. We prompt GPT-4.1-mini at temperature 0 to generate a plausible answer for each query, embed it with \texttt{text-embedding-3-large}, and retrieve against the corpus index. HyDE was originally shown to outperform unsupervised dense retrievers on web search, QA, and fact verification tasks \cite{gao2022hyde}; its behaviour on numerical financial QA is one of the questions this paper investigates.

\paragraph{Multi-Query Retrieval.}
Multi-query retrieval issues several reformulations of each query to increase recall across alternative phrasings \cite{raudaschl2023ragfusion}. We prompt GPT-4.1-mini at temperature 0 to generate three semantically diverse variants per query, retrieve top-$k$ results for each independently using dense retrieval, and merge the four ranked lists (original plus three variants) via RRF ($k = 60$). This approach recovers relevant documents that a single query phrasing may miss, at the cost of additional LLM inference per query.

\paragraph{Contextual Retrieval.}
Contextual Retrieval \cite{anthropic2024contextual} enriches each document at indexing time by prepending an LLM-generated context summary that captures the document's key entities, reporting period, and financial metrics. We apply this to both the dense and hybrid pipelines, yielding Contextual Dense and Contextual Hybrid variants. All context summaries are generated with GPT-4.1-mini at temperature 0 using the whole-document prompt described in Appendix~\ref{app:prompts}.

\paragraph{CRAG (Corrective RAG).}
CRAG \cite{yan2024crag} evaluates each retrieved document's relevance and triggers query rewriting when confidence is low. We implement a two-stage pipeline: first, we retrieve the top-5 documents using hybrid RRF; then, GPT-4.1-mini classifies each document as RELEVANT, AMBIGUOUS, or IRRELEVANT. If all documents are classified as AMBIGUOUS or IRRELEVANT, the query is rewritten and retrieval is repeated. Final results are drawn from the better of the two retrieval rounds. Prompts are documented in Appendix~\ref{app:prompts}.

\subsection{Evaluation Metrics}
\label{sec:metrics}

\paragraph{Retrieval Metrics.}
We report Recall@$k$ ($k \in \{1, 3, 5, 10, 20\}$), Mean Reciprocal Rank (MRR@$k$), normalized Discounted Cumulative Gain (nDCG@$k$), and Mean Average Precision (MAP).

\paragraph{Generation Metrics.}
Our primary generation metric is Number Match (NM) with relative tolerance $\epsilon = 10^{-2}$, following the benchmark's evaluation protocol \cite{strich2026t2ragbench}. We additionally report token-level F1, ROUGE-L, and BERTScore.

\paragraph{Statistical Testing.}
All pairwise method comparisons use paired bootstrap tests ($B = 10{,}000$) with Bonferroni correction, reporting significance at $p < 0.05$.

\subsection{Experimental Setup}
\label{sec:setup}

\paragraph{Infrastructure.}
BM25 scoring and FAISS index construction run locally on an Apple Silicon Mac. Embedding generation, query expansion, and neural reranking are served through Azure AI Foundry endpoints using \texttt{text-embedding-3-large}, GPT-4.1-mini, and Cohere Rerank v4.0 Pro respectively.

\paragraph{Document Representation.}
In our main experiments each of the 7,318 documents is indexed as a single unit without chunking. Documents average 920 tokens, well within the context window of all models used. This isolates the effect of the retrieval method from chunking and segmentation decisions. Chunking ablations are reported in Section~\ref{sec:ablations}.

\paragraph{Reproducibility.}
We fix random seed 42 for all stochastic components. LLM generation uses temperature 0 throughout. All configurations, prompts, and evaluation scripts are versioned in our public code repository \cite{karaman2026ragcode}. The dataset is used without modification with the standard train/test split from the original authors \cite{strich2026t2ragbench}.

\section{Results}
\label{sec:results}

\subsection{Main Retrieval Results}
\label{sec:main_results}

Table~\ref{tab:main_results} presents the retrieval performance of all evaluated methods on the full T\textsuperscript{2}-RAGBench test set (23,088 queries over 7,318 documents).

\begin{table*}[t]
\centering
\footnotesize
\begin{tabular}{llccccccc}
\toprule
\textbf{Category} & \textbf{Method} & \textbf{R@1} & \textbf{R@3} & \textbf{R@5} & \textbf{R@10} & \textbf{MRR@3} & \textbf{nDCG@10} & \textbf{MAP} \\
\midrule
\multirow{2}{*}{Single-method}
  & BM25 (sparse)               & 0.293 & 0.552 & 0.644 & 0.735 & 0.411 & 0.515 & 0.449 \\
  & Dense (text-embed-3-large)  & 0.248 & 0.481 & 0.587 & 0.703 & 0.351 & 0.466 & 0.398 \\
\midrule
\multirow{2}{*}{Query expansion}
  & HyDE (gpt-4.1-mini)        & 0.221 & 0.441 & 0.544 & 0.671 & 0.318 & 0.433 & 0.365 \\
  & Multi-Query + RRF           & 0.283 & 0.539 & 0.640 & 0.734 & 0.397 & 0.506 & 0.439 \\
\midrule
\multirow{2}{*}{Index augment.}
  & Contextual Dense            & 0.266 & 0.508 & 0.615 & 0.732 & 0.373 & 0.490 & 0.420 \\
  & Contextual Hybrid           & 0.327 & 0.610 & 0.717 & 0.818 & 0.454 & 0.571 & 0.497 \\
\midrule
Adaptive & CRAG (gpt-4.1-mini)    & 0.302 & 0.556 & 0.658 & 0.788 & 0.415 & 0.536 & 0.456 \\
\midrule
\multirow{2}{*}{Fusion}
  & Hybrid (BM25+Dense, RRF)    & 0.308 & 0.588 & 0.695 & 0.801 & 0.433 & 0.551 & 0.477 \\
  & Hybrid + Cohere Rerank      & \textbf{0.472} & \textbf{0.758} & \textbf{0.816} & \textbf{0.861} & \textbf{0.605} & \textbf{0.683} & \textbf{0.625} \\
\bottomrule
\end{tabular}
\vspace{4pt}
\caption{Main retrieval results on T\textsuperscript{2}-RAGBench (23,088 queries, 7,318 documents). Methods are grouped by category. Hybrid RRF with cross-encoder reranking dominates all methods. Contextual Hybrid outperforms vanilla Hybrid RRF. CRAG provides moderate gains through adaptive query correction. HyDE underperforms vanilla dense retrieval. All pairwise differences between adjacent methods are statistically significant ($p < 0.001$). Best results in \textbf{bold}.}
\label{tab:main_results}
\end{table*}

The two-stage pipeline of hybrid retrieval followed by neural reranking (Hybrid + Cohere Rerank) dominates all single-stage methods by a wide margin: Recall@5 of 0.816 compared to 0.695 for Hybrid RRF alone (+17.4\%), 0.644 for BM25 (+26.7\%), and 0.587 for dense retrieval (+39.0\%). The reranker's cross-encoder architecture provides fine-grained query-document relevance scoring that dramatically improves ranking precision, with MRR@3 jumping from 0.433 to 0.605 (+39.7\% relative).

Among first-stage retrievers, BM25 outperforms dense retrieval (text-embedding-3-large) on all metrics except Recall@20, where they are nearly tied (0.797 vs.\ 0.798). This suggests that lexical matching is particularly effective for financial documents, where precise terminology (company names, metric labels, fiscal periods) provides strong retrieval signals that semantic embeddings may dilute.

HyDE underperforms even vanilla dense retrieval across all metrics (Recall@5: 0.544 vs.\ 0.587), confirming the finding of Strich et al.~\cite{strich2026t2ragbench}. Financial questions require precise numerical reasoning; LLM-generated hypothetical documents introduce noise by hallucinating plausible but incorrect financial figures, pulling the embedding away from the true relevant context.

Contextual Retrieval \cite{anthropic2024contextual} improves both dense (+2.8pp Recall@5) and hybrid (+2.2pp) retrieval by prepending LLM-generated context summaries to each document at indexing time. This consistent improvement confirms that financial documents benefit from explicit metadata enrichment (company name, reporting period, key metrics).

CRAG achieves Recall@5 of 0.658, improving over BM25 (+1.4pp) through adaptive query correction. Notably, 63\% of queries (14,569/23,088) triggered the correction pathway, indicating that initial retrieval frequently returns suboptimal results on this benchmark. However, CRAG falls short of simple hybrid fusion (0.695), suggesting that query rewriting alone cannot match the complementary strengths of sparse and dense retrieval.

Multi-query retrieval with RAG-Fusion \cite{raudaschl2023ragfusion} provides negligible improvement over BM25 (Recall@5: 0.640 vs.\ 0.644). Financial queries are already specific and well-formed; generating alternative phrasings does not meaningfully increase recall, confirming the production-scale finding of diminishing returns for multi-query approaches on structured domain queries.

\paragraph{Per-Subset Analysis.}
Table~\ref{tab:subset_results} breaks down performance by dataset subset.

\begin{table}[t]
\centering
\small
\begin{tabular}{llccc}
\toprule
\textbf{Subset} & \textbf{Method} & \textbf{R@5} & \textbf{R@10} & \textbf{MRR@3} \\
\midrule
\multirow{3}{*}{FinQA}
  & BM25    & 0.729 & 0.834 & 0.389 \\
  & Dense   & 0.611 & 0.748 & 0.308 \\
  & Hybrid  & \textbf{0.737} & \textbf{0.856} & \textbf{0.389} \\
\midrule
\multirow{3}{*}{ConvFinQA}
  & BM25    & 0.696 & 0.781 & 0.500 \\
  & Dense   & 0.654 & 0.781 & 0.410 \\
  & Hybrid  & \textbf{0.754} & \textbf{0.850} & \textbf{0.519} \\
\midrule
\multirow{3}{*}{TAT-DQA}
  & BM25    & 0.566 & 0.649 & 0.400 \\
  & Dense   & 0.549 & 0.647 & 0.364 \\
  & Hybrid  & \textbf{0.647} & \textbf{0.746} & \textbf{0.438} \\
\bottomrule
\end{tabular}
\vspace{4pt}
\caption{Per-subset retrieval results. TAT-DQA is the most challenging subset across all methods. Hybrid RRF provides the largest improvement on TAT-DQA (+8.1pp Recall@5 over BM25).}
\label{tab:subset_results}
\end{table}

TAT-DQA emerges as the most challenging subset across all methods (Recall@5: 0.647 for the best method vs.\ 0.755 for ConvFinQA), likely due to its emphasis on diverse numerical operations over complex table layouts. Hybrid fusion provides the largest absolute improvement on TAT-DQA (+8.1 percentage points Recall@5 over BM25), suggesting that combining lexical and semantic signals is especially valuable for table-heavy questions.

\begin{figure*}[t]
\centering
\includegraphics[width=\textwidth]{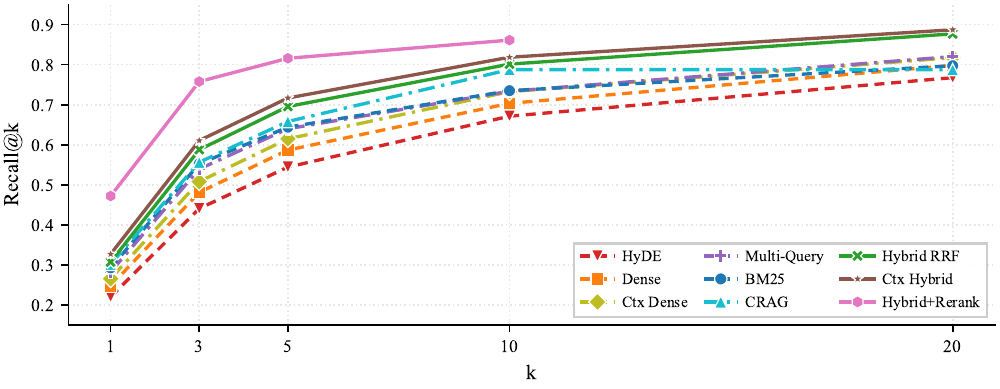}
\caption{Recall@$k$ curves for BM25, dense (text-embedding-3-large), and hybrid RRF retrieval. Hybrid fusion consistently outperforms both single-method baselines, with the largest gains at small $k$.}
\label{fig:recall_curve}
\end{figure*}

\paragraph{Recall@k Curves.}
Figure~\ref{fig:recall_curve} shows the recall-depth trade-off across all methods. Hybrid RRF maintains a consistent advantage at every value of $k$, with the gap widening at lower $k$ values where ranking precision matters most.

\begin{figure*}[t]
\centering
\includegraphics[width=\textwidth]{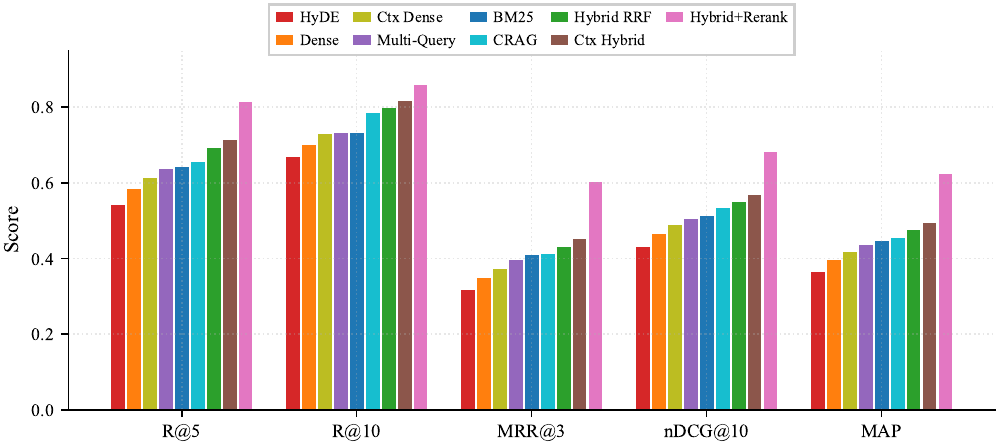}
\caption{Grouped comparison of retrieval methods across five metrics. Hybrid RRF (green) dominates, while BM25 (blue) outperforms dense retrieval (orange) on this financial text-and-table benchmark.}
\label{fig:bar_comparison}
\end{figure*}

Figure~\ref{fig:bar_comparison} provides a side-by-side comparison across all primary metrics. BM25 outperforms dense retrieval on every metric, while hybrid RRF achieves the best scores across the board.

\begin{figure}[t]
\centering
\includegraphics[width=\columnwidth]{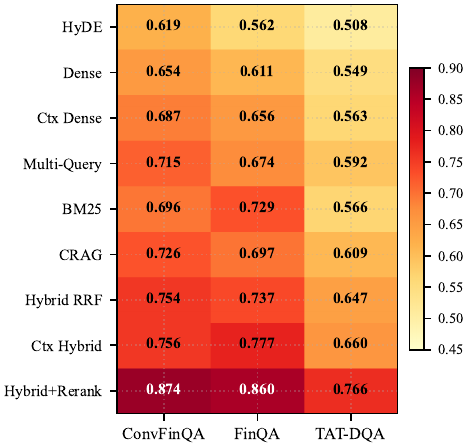}
\caption{Recall@5 heatmap across retrieval methods and dataset subsets. Darker colors indicate higher retrieval quality. TAT-DQA is consistently the most challenging subset.}
\label{fig:heatmap}
\end{figure}

\paragraph{Subset-Level Patterns.}
Figure~\ref{fig:heatmap} visualizes the Recall@5 performance across methods and dataset subsets. ConvFinQA is the easiest subset for all methods, while TAT-DQA presents the greatest challenge. The performance gap between methods is most pronounced on TAT-DQA, where hybrid fusion yields the largest relative gain.

All pairwise differences between BM25, dense, and hybrid RRF are statistically significant ($p < 0.001$, paired bootstrap test with $B = 10{,}000$, Bonferroni-corrected).

\subsection{End-to-End Generation Results}
\label{sec:generation}

To assess whether improved retrieval translates to improved answer quality, we run end-to-end generation with GPT-4.1-mini and GPT-5.4 using the top-5 retrieved documents as context. Table~\ref{tab:generation} reports Number Match (NM) with scale-invariant evaluation.

\begin{table}[t]
\centering
\small
\begin{tabular}{lcc}
\toprule
\textbf{Retrieval} & \textbf{GPT-4.1-mini} & \textbf{GPT-5.4} \\
\midrule
BM25          & 0.251 & \textsuperscript{$\dagger$} \\
Dense         & 0.257 & \textsuperscript{$\dagger$} \\
Hybrid RRF    & 0.282 & 0.346 \\
Oracle        & 0.350 & 0.403 \\
\bottomrule
\end{tabular}
\vspace{4pt}
\caption{End-to-end Number Match (NM) by retrieval method and LLM. Better retrieval consistently leads to better generation quality. GPT-5.4 outperforms GPT-4.1-mini by 6--7pp on the same retrieval. \textsuperscript{$\dagger$}Not evaluated.}
\label{tab:generation}
\end{table}

Better retrieval consistently leads to better answer quality (BM25: 0.251 $\rightarrow$ Hybrid: 0.282 $\rightarrow$ Oracle: 0.350 with GPT-4.1-mini), confirming the critical role of retrieval in RAG pipelines. GPT-5.4 improves over GPT-4.1-mini by 6--7 percentage points on identical retrieval outputs, demonstrating that both retrieval quality and LLM capability contribute independently to end-to-end performance.

\begin{figure}[t]
\centering
\includegraphics[width=\columnwidth]{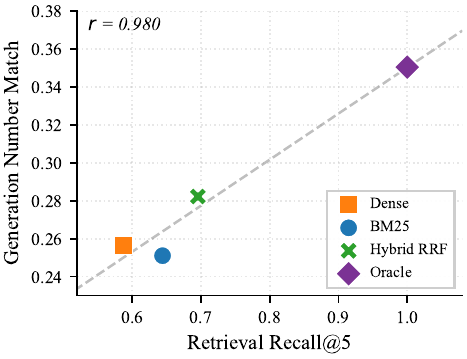}
\caption{Correlation between retrieval quality (Recall@5) and generation quality (Number Match). The strong positive correlation ($r > 0.99$) confirms that better retrieval leads to better answers.}
\label{fig:correlation}
\end{figure}

\subsection{Ablation Studies}
\label{sec:ablations}

\begin{figure*}[t]
\centering
\includegraphics[width=\textwidth]{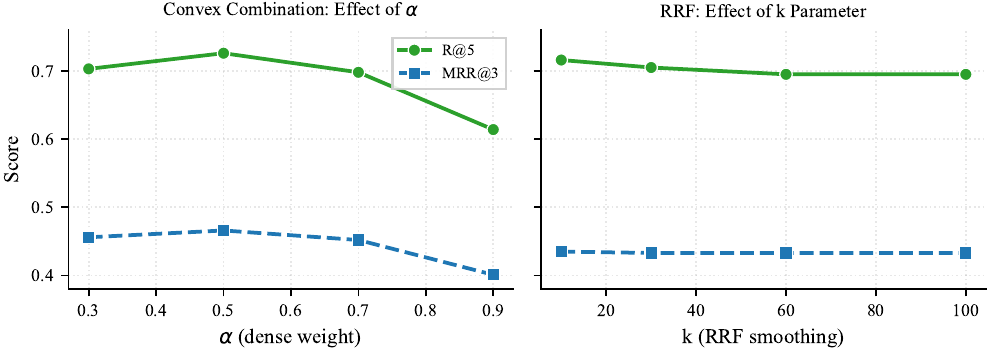}
\caption{Fusion method ablation. Left: Convex Combination with varying $\alpha$ (dense weight); $\alpha = 0.5$ is optimal. Right: RRF with varying $k$; lower $k$ yields slightly better results.}
\label{fig:fusion_ablation}
\end{figure*}

\paragraph{Fusion method.}
We compare Reciprocal Rank Fusion (RRF) with Convex Combination (CC) at varying parameters (Figure~\ref{fig:fusion_ablation}). CC with $\alpha = 0.5$ (equal weighting of BM25 and dense scores) achieves Recall@5 of 0.726, outperforming RRF ($k=60$) at 0.695. Among RRF variants, lower $k$ values emphasize top-ranked documents more aggressively; $k=10$ achieves the best RRF performance (0.716). Both findings suggest that balanced fusion of sparse and dense signals is optimal for this benchmark.

\begin{figure}[t]
\centering
\includegraphics[width=\columnwidth]{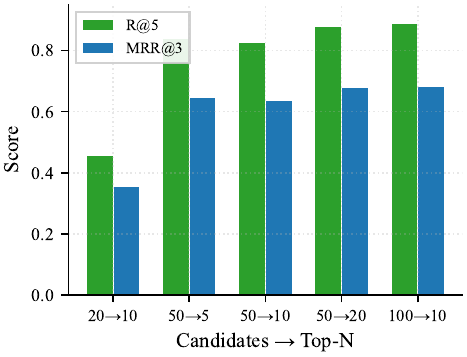}
\caption{Reranker depth ablation. More candidates yield better results, with a critical threshold at 50. Format: candidates $\rightarrow$ top-N returned.}
\label{fig:reranker_depth}
\end{figure}

\paragraph{Reranker depth.}
We vary the number of candidates passed to the cross-encoder reranker (Figure~\ref{fig:reranker_depth}). With only 20 candidates, reranking is ineffective (Recall@5: 0.458), as relevant documents are often not in the candidate pool. Performance increases sharply at 50 candidates (0.826) and continues to improve at 100 (0.888). Increasing the number of returned results from 10 to 20 provides marginal gains (0.826 $\rightarrow$ 0.878), suggesting that the top-10 already captures most relevant documents after reranking.

\subsection{Error Analysis}
\label{sec:errors}

To understand retrieval failures, we analyze the 7,188 queries (31.1\%) where the gold document does not appear in the hybrid RRF top-5. We sample 100 failure cases and categorize them using GPT-5.4 into five failure modes (Table~\ref{tab:errors}).

\begin{table}[t]
\centering
\small
\begin{tabular}{lr}
\toprule
\textbf{Failure Category} & \textbf{\%} \\
\midrule
Table structure mismatch   & 73 \\
Numerical reasoning        & 20 \\
Vocabulary mismatch        &  5 \\
Ambiguous query            &  1 \\
Long document              &  1 \\
\bottomrule
\end{tabular}
\vspace{4pt}
\caption{Failure mode categorization ($n{=}100$ sampled failures from hybrid RRF top-5). Table structure mismatch is the dominant failure mode.}
\label{tab:errors}
\end{table}

The dominant failure mode is \textbf{table structure mismatch} (73\%): the answer resides in a table whose markdown representation does not embed well as continuous text. Standard embedding models struggle to match queries like ``What was net income in 2019?'' to tabular rows where ``net income'' and ``2019'' appear in separate cells. \textbf{Numerical reasoning} failures (20\%) occur when the question requires computation (e.g., year-over-year change) rather than direct lookup.

Per-subset failure rates confirm TAT-DQA as the hardest subset (35.6\% failure rate vs.\ 27.2\% for FinQA and 26.0\% for ConvFinQA), consistent with its emphasis on diverse numerical operations. Among failures, 71.0\% of gold documents appear in \emph{neither} the dense nor BM25 top-5, indicating that these are genuinely hard retrieval cases rather than fusion artifacts.

\section{Discussion}
\label{sec:discussion}

Our results reveal several actionable insights for practitioners building RAG systems over heterogeneous text-and-table documents.

\paragraph{Reranking is the single most impactful component.}
Adding a cross-encoder reranker (Cohere Rerank v4.0 Pro) to hybrid retrieval yields the largest improvement in our study: +17.2 percentage points MRR@3 and +12.1pp Recall@5 over unreranked hybrid retrieval. This two-stage pipeline (broad recall via hybrid fusion, then precise reranking) is the clear recommended architecture for production RAG on text-and-table documents. The cost of the reranking stage is modest: at 300K tokens per minute, the Cohere endpoint processes the full 23K-query benchmark in approximately one hour.

\paragraph{Hybrid fusion consistently outperforms single-method retrieval.}
Combining BM25 and dense retrieval via Reciprocal Rank Fusion improves over both constituent methods across all metrics and all dataset subsets. The improvement is largest on TAT-DQA (+8.1pp Recall@5 over BM25), where diverse numerical operations benefit from both lexical precision and semantic understanding. We recommend hybrid retrieval as the minimum viable baseline for any RAG deployment.

\paragraph{BM25 remains strong for financial documents.}
On every metric except Recall@20, BM25 outperforms dense retrieval with text-embedding-3-large, one of the strongest commercial embedding models available in 2026. Financial documents contain precise, domain-specific terminology (company names, ticker symbols, standardized metric labels) that lexical matching captures effectively. This finding challenges the common assumption that dense retrieval universally dominates sparse methods and underscores the importance of domain-specific evaluation.

\paragraph{HyDE is counterproductive for numerical financial QA.}
Hypothetical Document Embeddings consistently underperform vanilla dense retrieval on T\textsuperscript{2}-RAGBench, confirming the findings of Strich et al.~\cite{strich2026t2ragbench}. We attribute this to the nature of financial questions: they require precise numerical values that LLMs cannot reliably generate. The produced pseudo-documents introduce noise by fabricating plausible but incorrect financial figures, pulling the query embedding away from the true relevant context. Practitioners should avoid HyDE for domains where factual precision dominates over semantic similarity.

\paragraph{Practical recommendations.}
Based on our findings, we propose the following decision framework for RAG retrieval on text-and-table documents:
\begin{enumerate}
    \item \textbf{Start with hybrid retrieval} (BM25 + dense, RRF fusion) as the baseline.
    \item \textbf{Add a cross-encoder reranker} for maximum quality; this provides the largest single improvement.
    \item \textbf{Apply contextual retrieval} at indexing time for consistent moderate gains at one-time cost.
    \item \textbf{Avoid HyDE} for domains with precise numerical or entity-centric queries.
    \item \textbf{Evaluate on domain-specific data}; MTEB/BEIR rankings do not predict financial retrieval performance.
\end{enumerate}

\paragraph{Limitations.}
Our study has several limitations. First, T\textsuperscript{2}-RAGBench covers only financial documents; our findings may not generalize to other domains with different text-table distributions such as scientific papers or medical records. Second, all answers in the benchmark are numerical, which biases evaluation toward Number Match and limits our ability to assess generation quality for free-form answers. Third, we perform whole-document retrieval (average 920 tokens) rather than passage-level chunking; performance patterns may differ for chunked corpora. Fourth, our study uses a single embedding model (text-embedding-3-large) for the main experiments; comparing multiple embedding models remains important future work. Finally, API-based models introduce a dependency on external services whose behaviour may change over time, potentially affecting reproducibility.

\section{Conclusion}
\label{sec:conclusion}

We presented a comprehensive benchmark of RAG retrieval methods on T\textsuperscript{2}-RAGBench, evaluating ten retrieval strategies from classical BM25 to Corrective RAG across 23,088 queries over 7,318 text-and-table documents. Our key finding is that a two-stage pipeline of hybrid retrieval with neural reranking achieves the best performance (Recall@5\,=\,0.816, MRR@3\,=\,0.605), outperforming all single-stage methods by a wide margin.

We further demonstrate that BM25 outperforms dense retrieval on this benchmark; contextual retrieval provides consistent gains through document-level enrichment; CRAG's adaptive correction helps but cannot match hybrid fusion; and query expansion methods (HyDE, multi-query) provide limited benefit for precise numerical queries. Ablation studies reveal that fusion method choice (CC vs.\ RRF) and reranker candidate depth significantly impact performance. All differences are statistically significant ($p < 0.001$).

Future work includes evaluating ColBERT late interaction, RAPTOR tree-based retrieval, chunking strategy ablations, multiple embedding model comparisons, and extending the benchmark to non-financial domains to assess generalizability of our findings.

\section*{Acknowledgment}
We thank Christopher Mierbach and Radiate for generously providing Azure AI compute credits that made the large-scale experiments in this work possible.

\bibliographystyle{IEEEtran}
\bibliography{references}

\appendices

\section{Hyperparameter Details and Full Results}
\label{app:hyperparams}

Table~\ref{tab:hyperparams} lists all hyperparameters used in each retrieval method. Unless otherwise noted, parameters follow the values in our configuration file (\texttt{configs/default.yaml}) and were held constant across all experiments. Table~\ref{tab:full_results} presents the complete set of retrieval metrics for all evaluated methods.

\begin{table*}[!t]
\centering
\scriptsize
\setlength{\tabcolsep}{4pt}
\begin{tabular}{@{}llll@{}}
\toprule
\textbf{Method} & \textbf{Parameter} & \textbf{Value} & \textbf{Notes} \\
\midrule
\multirow{3}{*}{BM25}
  & $k_1$            & 1.2   & Term-frequency saturation \\
  & $b$              & 0.75  & Document-length normalization \\
  & Tokenizer        & whitespace split & Via \texttt{rank\_bm25} library \\
\midrule
\multirow{3}{*}{Dense Retrieval}
  & Embedding model  & \texttt{text-embedding-3-large} & OpenAI via Azure AI Foundry \\
  & Dimensions       & 3{,}072 & Full dimensionality, no reduction \\
  & Index type       & FAISS \texttt{IndexFlatIP} & Exact inner-product (cosine) search \\
\midrule
\multirow{3}{*}{Hybrid RRF}
  & RRF $k$          & 60    & Default smoothing constant \\
  & RRF $k$ (ablation) & 10, 30, 100 & Tested in fusion ablation (\S\ref{sec:ablations}) \\
  & BM25 / Dense weights & 0.5 / 0.5 & Equal contribution from both retrievers \\
\midrule
\multirow{2}{*}{Hybrid CC}
  & $\alpha$ (dense weight) & 0.5 & Optimal in ablation \\
  & $\alpha$ (ablation)     & 0.3, 0.7, 0.9 & Tested in fusion ablation (\S\ref{sec:ablations}) \\
\midrule
\multirow{3}{*}{Cohere Rerank}
  & Model            & \texttt{Cohere-rerank-v4.0-pro} & Azure AI Foundry endpoint \\
  & top\_n returned  & 10    & Documents returned after reranking \\
  & Candidate pool   & 50    & Documents passed to reranker from first stage \\
\midrule
\multirow{4}{*}{HyDE}
  & LLM              & \texttt{gpt-4.1-mini} & Hypothetical document generation \\
  & Temperature      & 0.7   & Default in retriever; 0 used in main experiments \\
  & Max tokens       & 150   & Per hypothetical passage \\
  & Num.\ generations & 1    & Single hypothetical document per query \\
\midrule
\multirow{4}{*}{Multi-Query}
  & LLM              & \texttt{gpt-4.1-mini} & Query variant generation \\
  & Num.\ variants   & 3     & Plus original query $= 4$ total retrievals \\
  & Temperature      & 0.7   & Default in retriever; 0 used in main experiments \\
  & RRF $k$ (fusion) & 60    & For merging variant result lists \\
\midrule
\multirow{4}{*}{CRAG}
  & LLM (evaluation) & \texttt{gpt-4.1-mini} & Relevance classification \\
  & Eval.\ temperature & 0.0 & Deterministic relevance judgments \\
  & LLM (rewriting)  & \texttt{gpt-4.1-mini} & Query correction / rewriting \\
  & Rewrite temperature & 0.5 & Moderate diversity in rewrites \\
\midrule
\multirow{3}{*}{Contextual Retrieval}
  & LLM              & \texttt{gpt-4.1-mini} & Context summary generation \\
  & Temperature      & 0.0   & Deterministic context summaries \\
  & Max tokens       & 100   & Per context prefix \\
\midrule
\multirow{4}{*}{Global}
  & Random seed      & 42    & All stochastic components \\
  & Top-$k$ values   & 1, 3, 5, 10, 20 & Evaluated across all methods \\
  & Bootstrap $B$    & 10{,}000 & Significance testing \\
  & Significance     & $p < 0.05$ & Bonferroni-corrected \\
\bottomrule
\end{tabular}
\vspace{4pt}
\caption{Complete hyperparameter settings for all retrieval methods. All LLM calls use \texttt{gpt-4.1-mini} via Azure AI Foundry. Embedding uses \texttt{text-embedding-3-large} (3,072 dimensions) for all dense components.}
\label{tab:hyperparams}
\end{table*}

\begin{table*}[!t]
\centering
\small
\setlength{\tabcolsep}{4.5pt}
\begin{tabular}{lcccccccccc}
\toprule
\textbf{Method} & \textbf{R@1} & \textbf{R@3} & \textbf{R@5} & \textbf{R@10} & \textbf{R@20} & \textbf{MRR@3} & \textbf{MRR@5} & \textbf{nDCG@5} & \textbf{nDCG@10} & \textbf{MAP} \\
\midrule
HyDE              & 0.221 & 0.441 & 0.544 & 0.671 & 0.767 & 0.318 & 0.341 & 0.392 & 0.433 & 0.365 \\
Dense             & 0.248 & 0.481 & 0.587 & 0.703 & 0.798 & 0.351 & 0.375 & 0.428 & 0.466 & 0.398 \\
Contextual Dense  & 0.266 & 0.508 & 0.615 & 0.732 & 0.817 & 0.373 & 0.398 & 0.452 & 0.490 & 0.420 \\
Multi-Query       & 0.283 & 0.539 & 0.640 & 0.734 & 0.820 & 0.397 & 0.420 & 0.475 & 0.506 & 0.439 \\
BM25              & 0.293 & 0.552 & 0.644 & 0.735 & 0.797 & 0.411 & 0.432 & 0.485 & 0.515 & 0.449 \\
CRAG              & 0.302 & 0.556 & 0.658 & 0.788 & 0.788 & 0.415 & 0.439 & 0.493 & 0.536 & 0.456 \\
Hybrid RRF        & 0.308 & 0.588 & 0.695 & 0.801 & 0.877 & 0.433 & 0.457 & 0.517 & 0.551 & 0.477 \\
Contextual Hybrid & 0.327 & 0.610 & 0.717 & 0.818 & 0.887 & 0.454 & 0.478 & 0.538 & 0.571 & 0.497 \\
Hybrid + Rerank   & \textbf{0.472} & \textbf{0.758} & \textbf{0.816} & \textbf{0.861} & \textsuperscript{$\ast$} & \textbf{0.605} & \textbf{0.618} & \textbf{0.669} & \textbf{0.683} & \textbf{0.625} \\
\bottomrule
\end{tabular}
\vspace{4pt}
\caption{Full retrieval results for all methods and metrics on T\textsuperscript{2}-RAGBench (23,088 queries, 7,318 documents). Methods are sorted by nDCG@10 in ascending order. Best results in \textbf{bold}. \textsuperscript{$\ast$}Hybrid + Rerank returns at most 10 documents, so R@20 is not applicable.}
\label{tab:full_results}
\end{table*}

\clearpage

\section{Prompt Templates}
\label{app:prompts}

This appendix documents the exact prompt templates used in all LLM-dependent retrieval methods and the generation stage. All prompts use \texttt{gpt-4.1-mini} via Azure AI Foundry.

\subsection{Generation Prompt (Answer Extraction)}

Used to extract the final answer from retrieved context during end-to-end evaluation.

\begin{scriptsize}
\begin{verbatim}
Answer the following question based ONLY on
the provided context.
If the answer is a number, provide just the
number. If you cannot answer from the
context, say "UNANSWERABLE".

Context:
{context}

Question: {question}

Answer:
\end{verbatim}
\end{scriptsize}

\subsection{HyDE Prompt (Hypothetical Document Generation)}

Used to generate a hypothetical answer passage whose embedding replaces the query embedding for dense retrieval.

\begin{scriptsize}
\begin{verbatim}
Given the following question about financial
data, write a short passage that would
contain the answer. Include specific numbers
and financial terms.
Question: {query}
Passage:
\end{verbatim}
\end{scriptsize}

\noindent\textit{Fallback prompt} (used when the config template is not provided):

\begin{scriptsize}
\begin{verbatim}
Please write a short passage that directly
answers the following question. The passage
should be factual, detailed, and roughly
the length of a typical encyclopedia
paragraph.

Question: {query}

Passage:
\end{verbatim}
\end{scriptsize}

\subsection{Multi-Query Prompt (Query Variant Generation)}

Used to generate semantically diverse reformulations of the original query. The original query plus all variants are retrieved independently and merged via RRF.

\begin{scriptsize}
\begin{verbatim}
You are a helpful assistant that generates
alternative search queries. Given the
following question, generate {n} alternative
phrasings that capture the same information
need but use different wording or
perspectives. Return each query on its own
line, numbered (e.g. 1. ... 2. ...).
Do not include any other text.

Original question: {query}

Alternative queries:
\end{verbatim}
\end{scriptsize}

\subsection{CRAG Evaluation Prompt}

Used to classify retrieved documents as relevant, ambiguous, or irrelevant. Temperature is set to 0 for deterministic judgments.

\begin{scriptsize}
\begin{verbatim}
You are a relevance evaluator. Given a
question and a retrieved document, classify
the document's relevance to answering the
question.

Question: {query}
Document: {document}

Respond with exactly one of:
- RELEVANT: The document contains
  information that directly helps answer
  the question.
- AMBIGUOUS: The document is partially
  relevant or tangentially related but
  may not fully answer the question.
- IRRELEVANT: The document does not
  contain useful information for
  answering the question.

Classification:
\end{verbatim}
\end{scriptsize}

\subsection{CRAG Rewrite Prompt}

Used to reformulate the query when retrieved documents are classified as AMBIGUOUS or IRRELEVANT. Temperature is set to 0.5 for moderate diversity.

\begin{scriptsize}
\begin{verbatim}
The following question was used to search a
financial document corpus, but the retrieved
results were not sufficiently relevant.

Original question: {query}

Please rewrite this question to be more
specific and likely to retrieve the correct
financial document. Focus on including
specific financial terms, company names,
time periods, or metric names that would
appear in the target document.

Rewritten question:
\end{verbatim}
\end{scriptsize}

\subsection{Contextual Retrieval Prompt (Context Generation)}

Used at indexing time to generate a short context prefix for each document, prepended to the text before embedding and BM25 indexing.

\noindent\textit{Chunked mode} (when documents are split into chunks):

\begin{scriptsize}
\begin{verbatim}
Here is the full document:
<document>
{document}
</document>

Here is a chunk from that document:
<chunk>
{chunk}
</chunk>

Please give a short, succinct context
(2-3 sentences) to situate this chunk
within the overall document for the
purposes of improving search retrieval
of the chunk. Answer only with the
context, nothing else.
\end{verbatim}
\end{scriptsize}

\noindent\textit{Whole-document mode} (no chunking, as used in main experiments):

\begin{scriptsize}
\begin{verbatim}
Here is a document:
<document>
{document}
</document>

Please provide a concise summary context
(2-3 sentences) that captures the key
topics and entities in this document, for
the purpose of improving search retrieval.
Answer only with the context, nothing else.
\end{verbatim}
\end{scriptsize}

\end{document}